# Modeling-informed policy, policy evaluated by modeling

Evolution of mathematical epidemiology in the context of society and economy

Sitabhra Sinha[1,2,3]

1 The Institute of Mathematical Sciences, CIT Campus, Taramani, Chennai 600113, India
2 Center of Excellence in Complex Systems and Data Science (CoE-CSDS), IMSc
3 Homi Bhabha National Institute, Anushaktinagar, Mumbai 400 094, India

**Abstract**

The COronaVIrus Disease 2019 (COVID-19) pandemic that has had the world in its grip from the beginning of 2020, has resulted in an unprecedented level of public interest and media attention focusing on the field of mathematical epidemiology. Ever since the disease came to worldwide attention, numerous models with varying levels of sophistication have been proposed; many of these have tried to predict the course of the disease over different time-scales, ranging from attempting to project the number of active cases over successive days and weeks, to attempts at forecasting the date(s) on which subsequent ``waves'' of the pandemic will purportedly emerge. Other models have examined the efficacy of the various policy measures that have been adopted (including the unparalleled use of ``lockdowns'', i.e., extremely stringent restrictions on travel, social interaction, and access to public spaces) by countries around the world in an attempt to contain and combat the disease. This multiplicity of models may have given the impression of an apparent over-abundance of distinct mathematical approaches to investigate how pandemics evolve over time. More importantly, some of the more extravagant claims made by a few modeling groups about their ability to predict future outcomes, that went hand in hand with the occasional abuse of models by agencies having vested interests, have led to bewilderment in many quarters about the true capabilities and utility of mathematical modeling. Here we provide a brief guide to the epidemiological modeling enterprise, focusing on how it has emerged as a tool for informed public-health policy-making and has in turn, influenced the design of interventions aimed at preventing disease outbreaks from turning into raging epidemics. We show that the diversity of models is somewhat illusory, as the bulk of them are rooted in the compartmental modeling framework that we describe here. While its basic structure may appear to be a highly idealized description of the processes at work, we show that features that provide more realism, such as the community organization of populations or strategic decision-making by individuals, can be incorporated in such models to make them behave in accordance with empirical observations. We conclude with the argument that the true value of models lies in their ability to test *in silico* the consequences of different policy choices in the course of an epidemic, a much superior alternative to trial-and-error approaches that are highly costly in terms of both lives and socio-economic disruption.

**Introduction**

Every new pandemic brings about renewed interest in understanding the factors responsible for the rapid spread of diseases and the possible means by which they can be contained, and subsequently eliminated. Apart from spurring developments in research in pathogen biology and advances in medical practice for handling infections, the epidemics that broke out in the 20$^{th}$ and 21$^{st}$ centuries have brought to attention of public authorities the role that mathematical modeling can play in identifying key drivers of spreading, suggesting methods of breaking the chain of infection (Anderson and May ,1992; Keeling and Rohani, 2008). Modeling also helps in analyzing the effectiveness of possible remedial measures, which may be either pharmaceutical in nature, e.g., designing the most efficient strategy for vaccinating the population if a vaccine is available but can only be given to a limited number of people within a specific timeframe owing to resource and personnel constraints, or involve non-pharmaceutical means such as quarantining, travel restrictions, workplace and school closure, etc. Indeed, it is arguably a mathematical understanding of the herd immunity conferred upon vaccinating even a part of the population that made it possible to aim for complete elimination of a life-threatening disease, as was achieved for smallpox in 1980 when WHO declared it to have been eradicated from the entire world (Henderson, 1976; Henderson, 2011). Unfortunately, this success could not be emulated with other vaccine-preventable diseases as yet – although much of the world has been made polio-free through sustained vaccination campaigns (Larson and Ghinai, 2011). More alarmingly, childhood diseases such as measles for which effective vaccines have existed for decades are appearing to make a comeback even in affluent countries as a result of vaccine hesitancy (Hotez et al, 2020). As this potentially threatens reversing the great gains made in public health over the previous century through mass vaccination, mathematical modeling to identify the factors that may reduce vaccine uptake in the population serves a very important purpose in preventing future pandemics (Bauch and Bhattacharyya, 2012).

Thus, modeling has played multifarious roles in shaping epidemiological policy for several decades before the COVID-19 pandemic broke out in early 2020. However, what appears to be different this time as we struggle with the pandemic even after two years from the initial outbreak in Wuhan, China in December 2019, is the prominent part that modeling has played in driving the various policies adopted by countries around the world to contain further spread and reduce mortality (Hale et al, 2021). Unlike epidemics in the recent past such as the SARS outbreak in 2003 (see, e.g., Hufnagel et al, 2004, Chen et al., 2007) or the more widespread 2009 swine flu pandemic (Jesan et al, 2010), which were also occasions in which modeling was extensively employed, during this pandemic modeling (and modelers) has been prominently featured in the media and mathematical epidemiological terms such as R-nought have become common household terms. Moreover, results from model simulations have been cited by the authorities to support rolling out different measures or at least justify them afterwards. Of course, the real-time use of model simulations to understand the course of an ongoing epidemic and testing the efficacy of possible counter-measures which are then applied in the field is not in itself new. For example, during the 2001 epidemic of foot-and-mouth disease (FMD) among livestock in the United Kingdom, modeling was used to compare the effectiveness of country-wide pre-emptive vaccination against a policy that combined reactive vaccination and culling in affected farms (Ferguson et al, 2001). Also, detailed agent-based models taking into account information about the movements throughout the day of individuals residing in a city have already been used earlier to determine strategies that would be the most effective response to sudden outbreak of an infectious disease, possibly via deliberate release of pathogens within the urban environment by hostile forces (Barret et al, 2005).

The present pandemic has, however, dwarfed previous efforts at using models to devise policy by the sheer volume of literature that has been produced over such a short time, and the attention that has been given to various scenarios described by the models by the public, the media and governmental agencies is possibly unprecedented. It has highlighted as never before the role that data-driven quantitative analysis (and specifically, epidemiological modeling) can play in arriving at crucial decisions affecting public health (Van Kerkhove et al, 2010). It has also revealed the limitations of modeling when it comes to making accurate forecasts of a future outbreak, an inherently stochastic event which is affected by a multitude of factors not all of which are well-understood or for which data is available. Indeed, while models that involve a large number of variables and mechanisms can be brought forth to describe a pandemic, these typically also involve fitting many parameters from data that is inherently noisy and often incomplete – so that in practice, simpler models with fewer parameters may be more useful. Furthermore, modeling allows performing *gedankensperiments* whereby the impact of different policy decisions on the future progress of the pandemic can be gauged without having to go through a costly (both in terms of resources and lives lost) trial-and-error process. In this article, we first provide an outline of the early connection between modeling and public health policy design (indeed modeling itself arose from the need of providing convincing arguments for specific policies) before proceeding to describe how mathematical epidemiology has helped in designing vaccination policies that has been the cornerstone of the giant strides made in epidemic management from the later half of the last century. We go on to describe how developments in network theory has helped bring in a new age of data-informed public health policy design and finally, we show how the use of game-theoretic models are bringing understanding of the complex process by which individuals choose (or not) to adhere to the directives of public health authorities, which dictate the success or otherwise of such policies. We conclude with a short discussion of the future outlook based on the experience with managing the COVID19 pandemic.

**The Beginnings**

Surprisingly, the mathematical modeling of epidemics was pioneered not by a mathematician, but a medical doctor, Ronald Ross, who of course gained international renown for establishing the critical role that mosquitos play in transmitting malaria from one infected individual to another (for which he was awarded the second ever Nobel prize given for Physiology and Medicine in 1902). However, elucidation of the path of disease transmission was only the means towards an end for Ross, whose principal goal was the prevention of malaria. He continued to work towards this end with a single-minded focus, using the insights he had gained from his laboratory work to inform methods for controlling the spread of malaria (Nature, 1948). Ross had decided that controlling the population of mosquitos (the vector for malaria) is the key to this and in 1899 he was provided an opportunity to test his ideas for promoting public health when he headed a 3-member team sent in 1899 by the Liverpool School of Tropical Diseases to Sierra Leone, then a British colony in west Africa to investigate the causative factors of malaria and suggest means of controlling the spread of the disease (Bockarie et al, 1999). This was prompted by the extremely high mortality of the colonists infected by malaria, with the disease-induced death rate among European soldiers stationed in Sierra Leone being more than eight times higher than the soldiers recruited from the local population (Wilson, 1898). The team identified the two native mosquito species as being responsible for transmitting malaria to humans, a definitive demonstration that mosquito is responsible for human malaria[1] (Austen,1899). The team's recommendation for reducing the incidence of malaria involved antilarval measures such as applying tar to larger stagnant waterbodies and the removal of containers such bottles, cans or tyres – as well as filing in potholes in roads - which can function as storage for rainwater (and hence potential breeding ground for mosquitos). However, only limited funds were made available for these activities so that they could not be continued for any extended period of time. Not surprisingly, after the measures were stopped, mosquito breeding continued once more. Thus, as the measures did not appear to have had achieved a lasting reduction in malaria incidence, the authorities concluded that the results of the trial did not justify the expenses of carrying out a regular program for controlling mosquito breeding.

This first effort was subsequently followed by further experimental trials in malaria prevention by controlling mosquito breeding, such as by the US Army officers W. C. Gorgas and J. A. LePrince in Cuba in 1899 - which they followed up in collaboration with S.T. Darling in the Panama Canal Zone during 1904-14 (Gorgas, 1906) - and Ross himself in Ismailia, Egypt during 1901-1903 (Ross, 1903) and subsequently in Mauritius in 1908 (Ross, 1908). Ross had argued that the fact that malaria was widespread even in places where people used personal prophylaxis such as mosquito nets or treatment using quinine, indicated that these measures will not by themselves be effective in eradicating malaria unless accompanied by control of the population of mosquitos that carry the disease (Ross, 1903). He had pointed out that reduction of mosquito numbers by even a modest amount should have a substantial impact on the incidence of the disease. Rather than draining of marshes or large pools, this involved preventing mosquitos breeding in stagnant water that may have collected in much smaller tanks, or even pots and holes. Based on his experiences with authorities who were unwilling to continue devoting funds and manpower in a sustained manner into programs whose effectiveness they were unconvinced of[2], Ross realized that "proof" that such methods work will be difficult to provide based on field data alone. He therefore turned to modeling.

Ross had understood that for new cases of malaria to occur the parasite must be present in a large number of individuals and mosquitos that are capable of carrying it also will have to be numerous. If these conditions are not met, the chain of infection can be broken. This insight formed the basis for his public health program aimed at controling malaria, which he described using a simple quantitative argument (Ross, 1910). To begin with, Ross considered a single individual infected with Malaria in a population of 1000 individuals and asked what is the probability that a mosquito will be able to transmit the infection to another individual. Ross argued that as not every mosquito can bite a human, we can assume a 1 in 4 probability for such an event. Furthermore, such an event will have to involve the single infected individual, who is randomly chosen from the total village population of 1000, yielding a probability of 1/4000 that a mosquito will bite the infected person. As a mosquito does not immediately become infectious upon ingesting the parasite, we need to consider a probability (1 in 3, say) that it survives for the number of weeks required for the parasites to mature, and then factor in the probability that it will bite a human (1 in 4, as mentioned previously). Thus, the probability that a single mosquito will be able to infect another person from the index case is 1/48000. Thus, for the expected number of mosquitos that can continue the infection to be even of the order of unity, the population of the mosquitos will need to be at least tens of thousands. Thus, in order to control or eliminate malaria, one need not have to achieve complete extermination of mosquito, but bringing it down to levels that are practically achievable can potentially break further progress of the infection, so that once the infected person has recovered, there will be no further risk of malaria. This is a truly remarkable thought-experiment that shows how quantitative reasoning can help guide public health policy.

---

[1] In India, Ross had demonstrated the role of mosquito in transmitting avian malaria. The crucial step that showed human could contract malaria on being bitten by mosquitos that had previously bitten malaria-infected humans could not be experimentally established during this time because of a number of factors, including, perhaps understandably, the lack of volunteers willing to be bitten by infected mosquitos.

[2] As Ross put it in what he referred to as the sanitary axiom: "Widespread diseases … cause much pain, poverty, sorrow, expense and loss of prosperity … and the rule is to grudge spending a hundred pounds for disease which costs thousands…[Therefore] "for economic reasons alone, governments are justified in spending for the prevention of [malaria] a sum of money equal to the loss which the diseases inflict on the people" (Ross, 1910).

These arguments were put in a more formal mathematical form by Ross in 1911 by describing the evolution of the number of infected individuals in the population using a system of coupled differential equations (Ross, 1911). It is interesting that he framed this in terms of a very general process by which a certain event (which could be either, among other things, death or marriage or even bankruptcy) can occur to members of a population, and was not restricted to only the transmission of pathogens. Given a certain probability of such an event occurring to any one individual, Ross asked how many would be affected at any particular instant[3]. Between 1916-1917 in collaboration with the mathematician, Hilda Hudson, Ross developed a general theory of epidemics building upon this foundation. It was termed "a priori pathometry" by Ross, as instead of trying to understand *a posteriori* the reasons for the observed behavior of an epidemic behaved, the theory was aimed at describing the process by which an epidemic spreads in order to predict how such an event will develop over time (Ross 1916, Ross and Hudson 1917a, Ross and Hudson 1917b). To realize its full potential though, it had to be connected with parameters that could be measured from empirical data such that the output of the model could be made to quantitatively fit the field observations for a specific epidemic. This was done by George MacDonald in the 1950s when he revised the model for malaria spreading devised by Ross in order to include metrics for key components determining the nature of spreading by mosquitos, such as their death rate and feeding rate (Smith et al, 2012). This allowed one to extract results with important public health implications, most importantly, allowing identification of the most vulnerable link in the chain of transmission which could be targeted using control measures to arrest further spread of a disease.

---

[3] This "theory of happenings" of Ross can, thus, be seen as a precursor of attempts to understand social and economic behavior in terms of how events occurring to an individual can often be seen as an outcome of her interactions with other members of the population, an endeavor that has in recent times been often referred to as econophysics or sociophysics (Sinha et al, 2011).

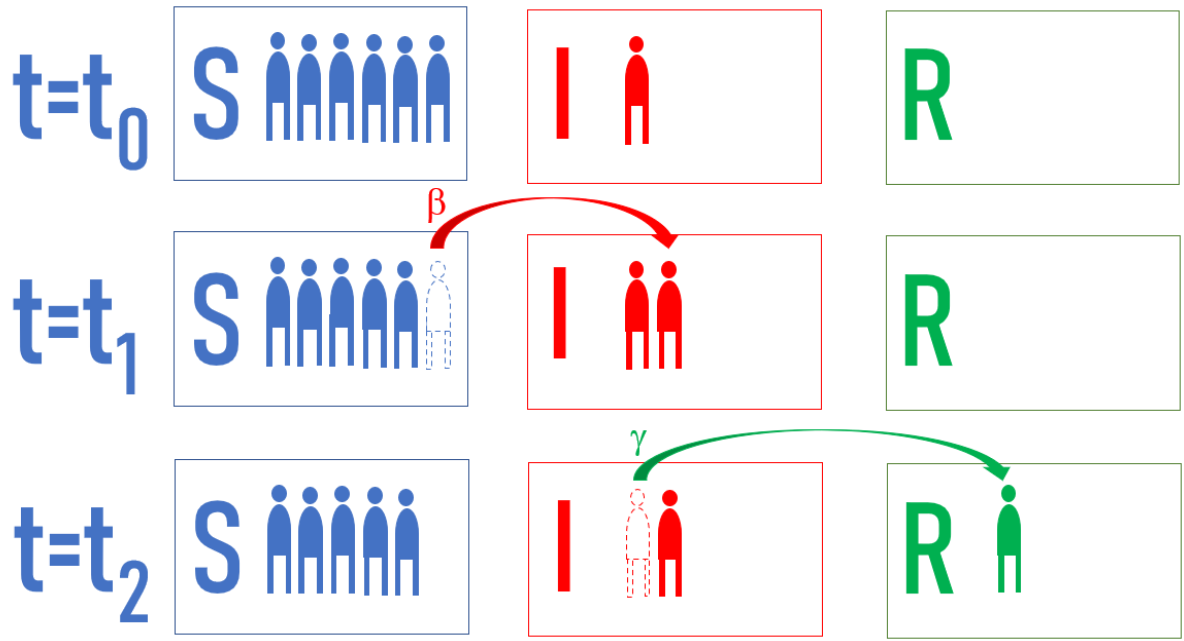

**Figure 1.** Schematic diagram of a Susceptible-Infected-Recovered compartmental model for epidemic progression. The rows represent the situation at successive time periods, with t=$t_0$ corresponding to the initial state when the first infections appear, with almost the entire population being susceptible. At subsequent times susceptible individuals move to the infected category and infected individuals move to the recovered category. The corresponding rates are indicated on the arrows representing these inter-compartment transfers.

**Compartmental Models of Epidemics**

While Ross's model is considered to be a pioneering effort quantitatively describing an epidemic, arguably the work of his protégé Anderson McKendrick (a military doctor, McKendrick had accompanied Ross to Sierra Leone and had been influenced by Ross's mathematical arguments) done in collaboration with William Kermack, a chemist by training, has been more influential in the development of mathematical epidemiology to its present stage (Kermack and McKendrick, 1927). While Ross had focused on the phenomenon of transmission of infection, Kermack and McKendrick considered the entire course of an epidemic, in particular, asking why and how does it end. In trying to answer this question, they developed a mathematical model for describing the dynamics of epidemics in general (i.e., not limited to a specific disease such as malaria) and showed that depending on the values of certain key parameters, an outbreak can either quickly die out spontaneously or result in an epidemic – thus pointing to the existence of an epidemic threshold. Furthermore, their model could explain why an epidemic dies out even when a part of the population is yet to be infected and the pathogen has not lost any of its ability to infect. For simplicity, they assumed the population to be *well-mixed*, in the sense, that any given individual could be in contact with any other individual at any time. This allowed them to approximate the expected number of new infections at a particular instant to be proportional to the number of infectious individuals, as well as the number of those who they can infect upon coming in contact (this latter category is termed as susceptible). Taking into account that those who have already been infected and have subsequently recovered develop a resistance or immunity to further infection, allows partitioning the population into a finite number of *compartments*, viz., those comprising individuals who are susceptible (S), infected (I) or recovered (R) (see Fig. 1).

The resultant SIR model is the simplest of the compartmental models, which sees the progress of the epidemic as transfer of part of the population successively from S to I to R compartments. The probability that any individual in S compartment will get infected (and hence be moved to the I compartment) upon coming in contact with an infected individual for time interval dt is given by β dt, where β is the rate of spreading. On the other hand, an infected individual can recover with a rate $\gamma$, and this recovery rate determines the rapidity with which individuals move from compartment I to compartment R. Under the well-mixed assumption, if i be the fraction of the population that is infected, then an individual having k contacts is likely to find ki of these to be with individuals who are infected. As each of these contacts has a probability β dt of successfully transmitting infection, the probability that at least one of these gives rise to a new infection is 1 – (1- β dt)$^{ki}$, which in the limit of extremely small time interval such that β dt <<1 simplifies to βki dt (retaining only the leading order in the power series expansion). If we also approximate the number of contacts k of each individual with its mean value ⟨k⟩ across the entire population, the resulting time-evolution equations for the fraction of population s,i,r residing in the compartments S,I,R respectively, are (Barrat et al, 2008):

ds/dt = −β ⟨k⟩ i s,

di /dt = β ⟨k⟩ i s − $\gamma$ i,

dr /dt = $\gamma$ i.

It has been implicitly assumed that the epidemic occurs at a much faster time-scale compared to that at which demographic changes take place, such that the total population summed over the compartments is constant, i.e., s+i+r=1. Thus, only two out of the three variables are independent. As by definition, an epidemic is said to occur when di/dt > 0 (i.e., the infected population grows over time), the threshold condition needed to be satisfied for an epidemic to take place is β ⟨k⟩ s − $\gamma$ > 0. Noting that the recovery rate is the reciprocal of the average period $\tau$ for which an individual is infectious, we can rewrite the condition as

β ⟨k⟩ $\tau$ s > 1      (1)

For a disease against which no vaccine is available, at the time of an initial outbreak almost the entire population can be assumed to be susceptible such that s ≈ 1. Under this circumstance, the infected fraction will evolve as di /dt = (β ⟨k⟩ − $\gamma$ ) i. Thus, if β ⟨k⟩ > $\gamma$, the fraction of infected will increase exponentially; conversely, it will rapidly decay to zero if β ⟨k⟩ < $\gamma$. As the solution of the equation provides a quantitative measure of the rate of growth of the infected population, it can be connected to one of the most important metrics for an epidemic, viz., the basic reproduction number ($R_0$). This measures the mean number of infections that result from contacts with a single infected individual in a wholly susceptible population (Heesterbeek, 2002), and can be expressed in terms of the parameters of the SIR model as $R_0$ = β ⟨k⟩ / $\gamma$.

It is easy to see from Eq (1) that the rate of progress of an epidemic is essentially influenced by four different factors, viz., (i) the ease with which a contact between infected and susceptible individuals result in a successful transmission of infection (β), (ii) the mean number of contacts between individuals (⟨k⟩) which is an attribute of the social network that they are part of, (iii) the duration for which a person is able to infect others ($\tau$), and (iv) the proportion of the population which is susceptible (s) (Anderson and May, 1992;Kucharski, 2020). As it is the product of these four quantities that yield the epidemic growth rate, it suggests a multi-pronged approach towards controlling an epidemic whereby any combination of these factors can be targeted by public health policy interventions to stop the epidemic. Thus, public hygienic practices such as use of masks and disinfection attempts to decrease β, while ⟨k⟩ is sought to be lowered by quarantining, school or workplace closure and physical distancing – and in extreme cases, by imposing shelter-in-place or lockdown. Measures put in place for rapid detection through mass testing and isolation of the identified infectious individuals aims at reducing τ. Finally, if a vaccine is (or becomes available), it can reduce s by transferring individuals from the S directly to the R partition, without having to enter the intervening I compartment. As some of the factors may be more amenable to reduction through the means available at hand, the usefulness of the expression of the growth rate in this product form is enormous, allowing us to determine by how much the achievable reduction in the different factors can slow down the spread of an epidemic. In this context, it is important to note that even small reductions in each of the factors can result in a relatively large decrease in the growth rate, possibly bringing it below the epidemic threshold. Thus, the public health policy implications of even such a simplified model are far-reaching.

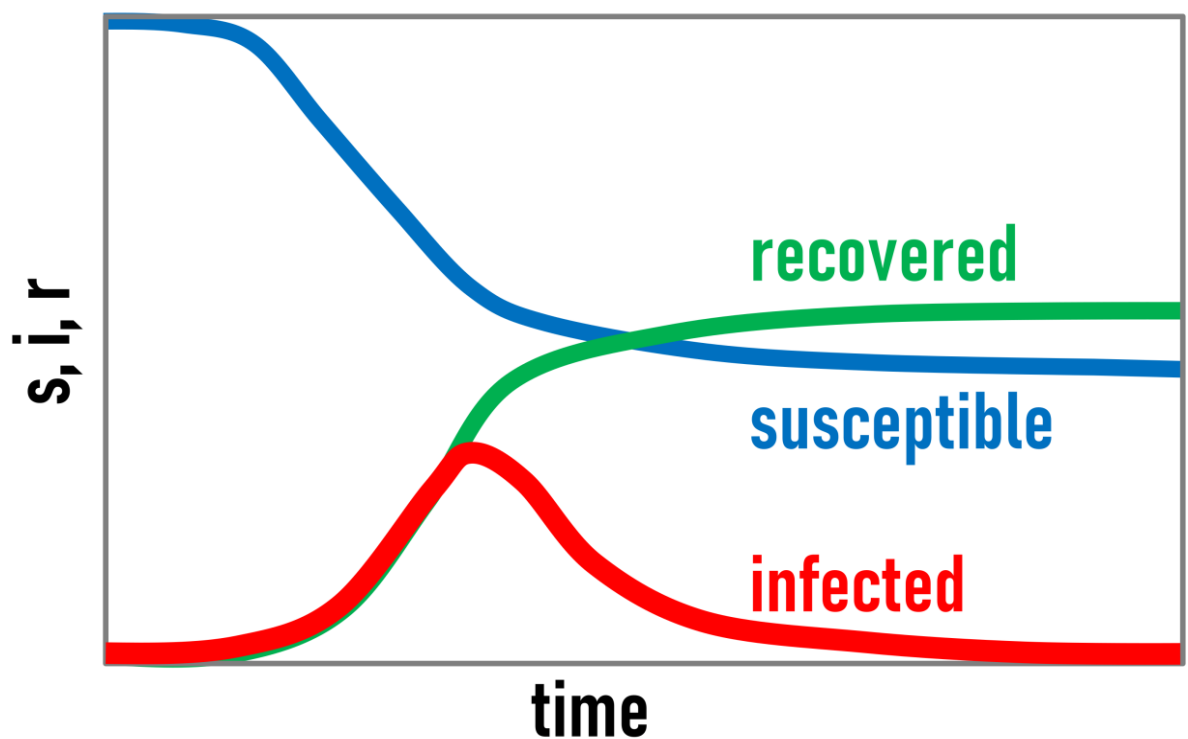

**Figure 2.** Typical time-evolution (shown schematically) of the susceptible, infected and recovered population fractions during a simulated epidemic in the SIR model.

Moving our focus away from the beginning of an epidemic, if we now consider its entire trajectory, we observe that the initial exponential rise of i eventually slows down, plateaus and then declines (Fig. 2). This is entirely consistent with the fact that the total population is conserved, and with time the number of individuals remaining in susceptible compartment will decline. Thus, after the epidemic has continued for some time, an infectious agent will mostly encounter individuals who are either already infected or have recovered from it (and are therefore immune). While there may still be susceptible individuals remaining, if they are screened from the pathogen by the majority that are in I or R compartments, the number of new successful transmissions will be few and far between. Eventually, when the pathogen can no longer find a susceptible individual over the time period during which all remaining infected individuals recover, the chain of transmission will be broken and the epidemic will come to an end even though there will still be susceptible individuals remaining in the population (see Fig. 2). This situation is usually referred to as one in which the population has reached *herd immunity*, such that even though there may be individuals who may still acquire the infection, collectively the population is able to resist any epidemic.

The size of the susceptible population remaining at the end of an epidemic turns out to be a function of the basic reproduction number, and is given by the implicit equation $s(\infty)=\exp(-R_0 [1-s(\infty)])$ (Diekmann and Heesterbeek, 2000). Its complement, i.e., $1-s(\infty)$, yields the final size of the epidemic, i.e., the total fraction of the population who would have been infected at one time or another before the epidemic runs its course, *if no measures are taken to mitigate it*. Needless to say, this is also an important metric from the perspective of public health. In the context of COVID19 epidemic, for which $R_0$ has been calculated to be between 2 and 3 for most countries, this translates to about 80%-94% of the population being eventually infected before herd immunity is achieved[4]. This has obvious bearing on discussions about the nature of measures that ought to be adopted to counter the epidemic. During the early stage of the COVID19 pandemic in 2020, several countries were debating whether it would be more beneficial to not introduce socio-economically costly lockdowns and just allow the population to achieve herd immunity by having the infection spread, albeit in a controlled manner (Khalife and VanGennep, 2021). Among advanced economies, the United Kingdom (to some extent) and most notably, Sweden, chose this path, arguing that by shielding the most vulnerable segment of the population from getting infected and assuming that the younger population, despite getting infected in large numbers, will be less likely to suffer life-threatening manifestation of the disease, the epidemic can be managed with minimum disruption. However, in practice, the policy not only led to more infections, higher level of hospitalization and greater number of deaths than that seen in the neighboring countries of Norway, Denmark and Finland which had adopted more interventionist approaches (Orlowski and Goldsmith, 2020). More worryingly, high levels of mortality persisted in Sweden for longer than its neighbors, running counter to the assumption underlying its strategy, i.e., trying to ``flatten the curve'' by adopting stricter measures to prevent infection would only prolong the epidemic, whereas allowing the disease to spread among the population would result in herd immunity being quickly achieved (thereby protecting the vulnerable population from infection).

While our discussion of compartmental modeling framework has only considered in detail the basic SIR model, one should point out that the framework is general enough to accommodate additional compartments that may be necessary to adequately describe the specific features of the progression of the disease whose epidemic is being investigated. Indeed, it appears that for describing the COVID19 epidemic it is important to take into account the latency period between transmission of the pathogen to a susceptible individual (exposure) and the pathogen starting to multiply within the individual (infection, which may be either symptomatic or asymptomatic). This is incorporated by inserting an additional compartment E, corresponding to the category of individuals who are exposed, between the S and I compartments of the model (which is thus termed as a SEIR model). Comparison between different models suggest that projections of disease incidence made using such SEIR models tend to be relatively more accurate (Friedman et al, 2020), and they have indeed been used widely to fit the evolution of the COVID19 pandemic in different countries and regions (see, e.g., IHME Covid 19 Forecasting Team, 2021 for forecasts of COVID19 scenarios in the United States of America).

---

[4] In reality, the exact value for the fraction of population that needs to be infected before reaching herd immunity COVID19 is hard to calculate owing to a number of factors, which has led to many different estimates being publicised at different times (McNeil, 2020). Some of these factors complicating the estimation include the different mitigation measures that have been undertaken at various stages, variation in the degree of susceptibility among individuals, presence of several strains of the virus to which a population may exhibit different resistances, and the possibility of re-infection among those who have previously been infected.

It is of course possible to introduce other compartments into an epidemic model, e.g., distinguishing between symptomatic and asymptomatic infected individuals. This may be an important distinction to include in the model if one is aiming to study the role that rapid identification and subsequent isolation of infectious individuals may play in containing the epidemic. While it may be difficult to identify asymptomatic cases - who may be themselves unaware that they are infected and will thus, not seek medical attention – in the absence of wide-spread (and repeated) testing, and yet these individuals are often just as likely to transmit the infection to others as symptomatic cases (Gao et al, 2021). It has been estimated that more than a third of all COVID19 infection cases are asymptomatic (Sah et al, 2021), which could be silently contributing to the chain of propagation of the infection, making surveillance-based control of the pandemic particularly challenging. Indeed, it has been reported that more than half of all COVID19 infections could have been caused by individuals not showing any overt signs of being infected, i.e., who were either presymptomatic (i.e., yet to show clinical symptoms) or asymptomatic (Moghadas et al, 2020)[5].

Other additional compartments may be included by further dividing the infected compartment into those who developed only mild symptoms and those for whom the disease manifested in an extremely severe form, which would be necessary if one is using the model to project hospital occupancy or ICU requirement. Again, to estimate death arising due to the infection, the R compartment may be divided into sub-compartments corresponding to those who recovered and patients who were removed from the population through death. The possibility of re-infection among those who have already been infected earlier (Stokel-Walker, 2021) can also be investigated in this modeling framework by assuming that individuals in the recovered category may with a certain rate of immunity loss re-enter the susceptible category, thereby allowing them to be infected once again. Such SIRS dynamics, when simulated on a social contact network, can exhibit persistently recurring outbreaks resembling multiple ``waves'' of epidemics (Jesan et al, 2016). However, one must keep in mind that increasing the complexity of the model by introducing many compartments does not necessarily lead to more accurate description of the actual epidemic. This is because of the much larger number of parameters governing the dynamics of these more complex models, each of which need to be fitted from the data and are hence susceptible to problems associated with parameter estimation (Basu and Andrews, 2013). Therefore, in practice, a simpler model having fewer compartments may be easier to calibrate with empirical data (and in a more robust manner), and hence be capable of providing a more accurate picture of an epidemic than one which has many more compartments in order to conform more closely with reality. Ultimately, which model is best for one's purpose, depends on the questions that are being asked.

---

[5] Distinguishing between presymptomatic and asymptomatic individuals may also be necessary in some cases. This can possibly be implemented by having the flow from the exposed (E) compartment bifurcate into two streams, one going over to asymptomatic compartment while the other successively passes through presymptomatic and symptomatic comparments, both eventually converging again in the R compartment.

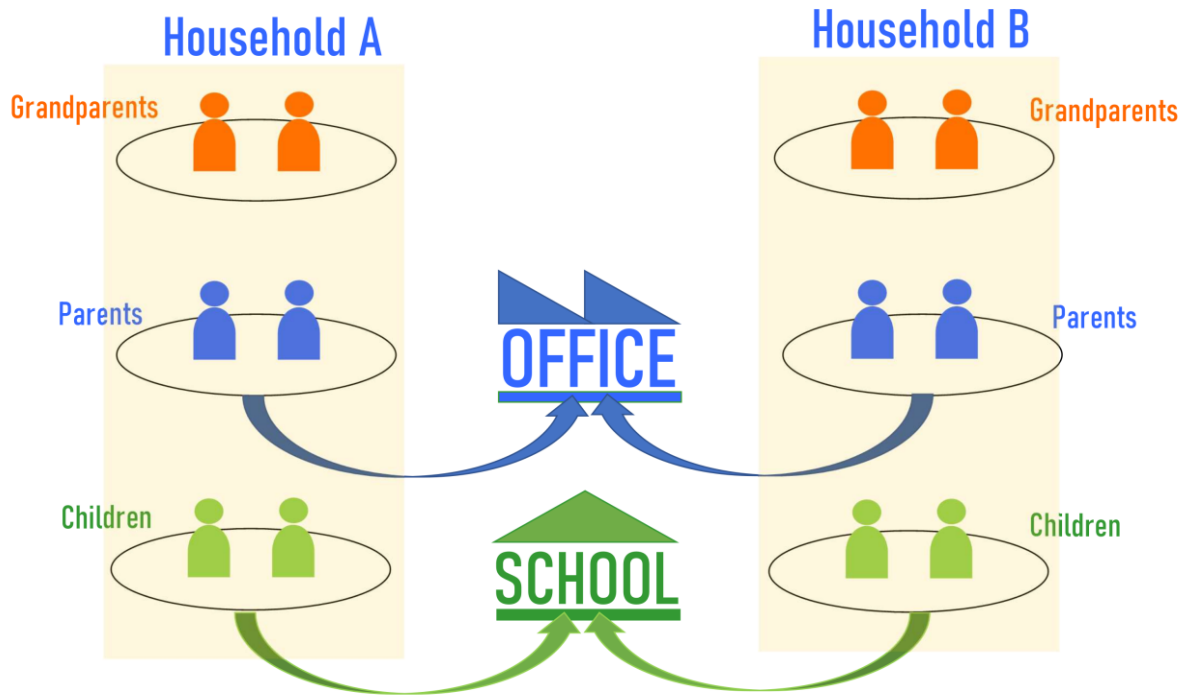

**Figure 3.** Considering the age-structure of a population is important in analyzing the spread of an epidemic not only because the effect of the disease on different age groups may differ drastically but also as the frequency of contacts (resulting in potential exposure) may differ remarkably between, as well as, across individuals belonging to different age groups. Within the same household, senior citizens may have much more limited contact with individuals outside their home, compared to adults or children, who will be in touch with others at offices and schools respectively.

The compartments could be further sub-divided according to innate characteristics of individuals (e.g., their age), if the resulting sub-populations differ in terms of the various rates that govern the epidemiological dynamics. Given the differential effect of COVID19 according to age, with older patients forming the bulk of the severe cases with increased mortality risk (Davies et al, 2020), several models, including one modeling the epidemic in India (Hazra et al, 2021) have considered age-structure of the population. The mobility pattern and contact rates of individuals in different age groups can be considerably different (Fig. 3), and are affected differently by targeted containment strategies such as school closure or workplace closure. Thus, it is expected that by considering the different rates of transmission within and between distinct age-categories, a more accurate picture of disease progression will be obtained (Wilder et al, 2020). Indeed, such studies suggest specific policies as regards optimal containment, taking into account economic impact of such policies. For example, it has been suggested that making half of the population of a particular age group shelter in place (i.e., undergo lockdown) while ensuring only physical distancing for the rest can substantially reduce both infections and deaths (Wilder et al, 2020). As this will have a much more limited social and economic cost compared to a complete lockdown it does seem a welcome alternative to the global lockdown policies followed by many governments around the world. In particular, modeling the spread of COVID19 on a simulated social network has led to a suggestion of implementing alternating quarantine procedure, whereby at any time, half the workforce works for one week while the other half shelters in place, with the roles being exchanged on alternate weeks (Meidan et al, 2021).

**Vaccination**

The idea of herd immunity for preventing an epidemic, originally conceived in terms of the critical fraction of the population that needs to be resistant against contracting the disease by prior infection, also applies when vaccination can be performed to make individuals immune without having them go through the infected stage (i.e., direct passage from the S to the R compartment). Assuming that the vaccine has a 100% effectiveness (i.e., anyone receiving it is completely immune and will never get infected), then if p fraction of the population is vaccinated, the fraction of the population that is still susceptible to the disease is s= 1 – p. From the condition for an epidemic to occur (Eq. 1) it is easy to see that in order to prevent an epidemic we will need to vaccinate more than $p_c$ = 1- (1/$R_0$) fraction of the population. Thus, for a disease like COVID19 where $R_0$ is reported mostly between 2 and 3, more than 50-66% of the population needs to be vaccinated to prevent any local outbreaks from spreading. Note that, this is a lower bound as none of the vaccines that have been introduced so far against the disease has 100% effectiveness, ranging instead from 95% down to 67% (Evans and Jewell, 2021); moreover, their efficacy appears to decrease with time (Katella, 2021). If a vaccine is only partially able to confer protection, it is intuitively clear that the threshold percentage of population needed to be vaccinated will need to be higher. Indeed, if we assume that a vaccine prevents infection in only $\varepsilon$ fraction of the recipients (i.e., $\varepsilon$ is the vaccine effectiveness in the field – which may differ from the efficacy observed in clinical trials), then at least (1– 1/$R_0$)/$\varepsilon$ fraction of the total population must be vaccinated in order to keep an outbreak localized. This implies that if the effectiveness is less than (1– 1/$R_0$), then the epidemic cannot be terminated even after the entire population has received the vaccine (Fine et al, 2011). The important implication for this result is that any COVID19 vaccine that has less than 66% effectiveness may not be worth introducing for public use, an important input for taking decisions on vaccine policy.

A useful comparison may be made with the worldwide vaccination program for the eradication of smallpox that was carried out by the WHO during 1966-1977. As estimates of $R_0$ for smallpox vary between 3.5 and 6 (Gani and Leach, 2001), the strategy was to achieve in every country more than 80% coverage through mass vaccination. This was obviously difficult to achieve in poorer economies, as is also seen in the present for COVID19 vaccination[6]. However, in places where there was insufficient vaccine to achieve the critical vaccinated fraction, an alternative strategy involving identifying every new case of smallpox and then containing the infection by vaccinating all primary contacts, as well as, secondary contacts (i.e., the contacts of primary contacts), turned out to be successful in eradicating the disease even in areas where the coverage was substantially lower than the threshold (e.g., in Nigeria it was less than 50%) (Belongia and Naleway, 2003). Indeed, such ring vaccination became the dominant strategy in the latter stages of the WHO smallpox eradication campaign. While this may suggest the use of similar methods for countering the poor coverage of COVID19 vaccines in several countries, it has to be noted that the strategy worked so well for smallpox even when there were delays in identifying cases partly because every infected person eventually manifests clinical symptoms of the disease and partly because it is transmitted only after relatively prolonged contact with the infected person (Deen and von Seidlein, 2018). In contrast, a large number of COVID19 cases are asymptomatic even when they are infectious and the pathogen spreads fairly easily, e.g., via aerosols that can be inhaled even by persons not in close proximity of the infected individual (Jarvis, 2020).

---

[6] As of December 6, 2021, while 72% of the population in USA and Canada have received at least one dose of a vaccine against COVID19, this is only 11% for Africa. For comparison, India during the same time has had 35% of the population fully vaccinated and 59% have received at least one dose. Globally, 56.5% have received at least one dose (Holder, 2021).

However, there are other characteristic features of COVID19 that could be considered to design optimal vaccine strategies. For example, the differences observed in the epidemic dynamics in different age-classes can be exploited. This is not unique to COVID19, as such age-structure effects can be seen, for instance, in influenza for which modeling has shown that the most effective reduction is achieved by preferentially vaccinating school-age children and young adults as this is more likely to break chains of infection (Medlock and Galvani 2009). Although there are similarities between these two diseases, it turns out that this is not true for COVID19 as the success of a strategy also depends on the effectiveness of the vaccine being used (Bubar et al. 2021). If a vaccine has lower efficacy, the indirect benefits of vaccination in protecting unvaccinated individuals from being exposed to the pathogen would also be reduced. As in such case, onward transmission may continue even when younger, more mobile individuals are vaccinated, it may be more effective in terms of reducing mortality to vaccinate the older population who have a higher risk of suffering from severe forms of the disease (possibly leading to death), especially as, unlike influenza vaccines, there is little evidence of age-dependence in the efficacy of COVID vaccines. (Bubar et al. 2021).

**Networks**

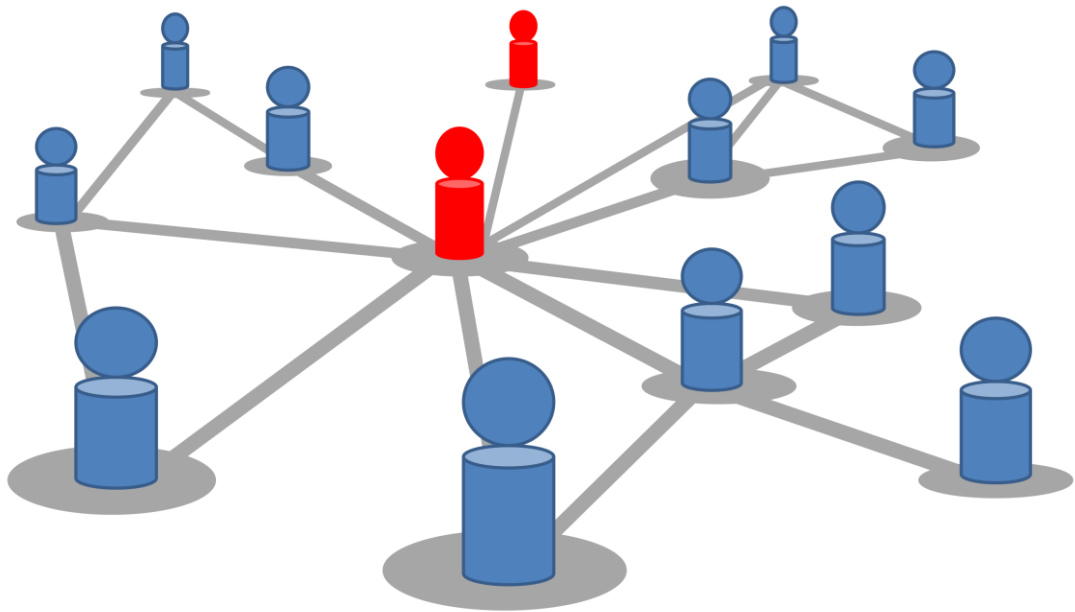

**Figure 4.** Social networks through which infections propagate from one individual to another are a more accurate reflection of the heterogeneity of contacts in a real population. The central node (indicate in red) with high *degree* (number of connections) can play a key role in the rapid spread of an epidemic through the population shown, once it gets infected by another node (also shown in red) that is otherwise peripheral to the network.

We have so far considered transmission in populations that are well-mixed (i.e., any pair of individuals in the population has the same probability of being in contact). This presumes a homogeneous contact structure, with every individual having more or less the same number of contacts that are randomly distributed among the population. While this may be interpreted in terms of random graphs, i.e., networks in which every pair of nodes has equal probability *p* of being connected, such a contact network structure does not accord with reality even though it makes the models mathematically tractable (Andersson and Britton, 2000). Developments in our understanding of the complex networks that underlie real-world systems (see Newman (2010) for a comprehensive review), has therefore also led to a better appreciation of how the structure of the connection topology affects progression of epidemics (Newman, 2002). In particular, many studies have focused on the role of degree heterogeneity, i.e., variations in the number of contacts that each individual has with others in the population (Barrat et al, 2008). This is particularly relevant for the spreading of COVID19 where the bulk of infections are believed to have been driven by relatively few super-spreading events[7] (Lee et al, 2020; Chen at al, 2021). Superspreading has also been implicated in the genesis of variants of concern of the SARS-Cov-2 pathogen (i.e., genetic mutants that have higher transmissibility or virulence compared to the dominant strain) by facilitating genetic drift (Gómez-Carballa et al, 2021). Modeling of epidemics in which superspreading plays a prominent role has suggested that in such processes most infections do not to give rise to sustained chains of transmissions and outbreaks are relatively uncommon, unlike in well-mixed models (Lloyd-Smith et al, 2005). However, a few rare events can lead to the cases amplifying to large numbers relatively quickly, as has indeed been borne out during the COVID19 pandemic (Wong and Collins, 2020). Given the high individual variation in transmissibility, the results suggest that measures targeting specific people or places for intervention may be more efficient in controlling the pandemic than instituting general population-wide measures (Lewis, 2021).

An intriguing possibility underlying the high variation of transmission ability that has been explore via modeling is the potential role of hubs in the contact network (Fig. 4), i.e., individuals having extremely large number of links compared to the average (Newman, 2010). A theoretical study on their effect in the COVID19 pandemic concluded that hubs result in increased spreading at least initially, but that they also tend to lower the threshold for attaining herd immunity (Großmann et al,2021). Intriguingly, for networks having heterogeneous degree distribution, it can be shown that the condition for the outbreak of an epidemic (instead of $\beta \langle k \rangle / \gamma > 1$ as in the case of homogeneous networks) is given by $\beta/\gamma > \langle k \rangle / (\langle k^2 \rangle - \langle k \rangle)$ (Barrat et al, 2008) . This result suggests that as the heterogeneity increases, epidemic threshold will be lowered (Pastor-Satorras et al 2015), making it easier for infections to develop into large outbreaks and also making them harder to prevent by vaccinations as much larger fraction of the population would need to be immunized against the disease. Even more intriguing possibilities arise if the network is scale-free[8], with the degree distribution decaying as a power law such that the second moment is diverging ($\langle k^2 \rangle \to \infty$). For networks of arbitrary size, a consideration appropriate for example in the context of computer viruses spreading over the internet, the condition implies that there are no thresholds to constrain epidemics from occurring (Pastor-Satorras and Vespignani, 2001). In other words, regardless of the fraction of the population made resistant to infection, herd immunity will never be achieved. While alarming, it has also been pointed out that as individuals rarely have close contact during the relatively brief period in which they are infectious with an extremely large number of people, the results may not apply to transmission of diseases such as SARS-Cov-2. On the other hand, it has ramifications for sexually transmitted diseases such as HIV which may reside in individuals for a long time without apparent symptoms, and spread through the activity of a few highly promiscuous individuals (Lloyd and May, 2001).

---

[7] For example, a study based on contact-tracing data during the SARS-Cov-2 outbreak in Wuhan, China, between January and April, 2020, concluded that as many as 80% of the cases could be attributed to only 15% of the infected individuals (Sun et al, 2021). This is consistent with findings from a study of COVID19 outbreak in several counties of Georgia, USA, that the top 2% (in terms of spreading the epidemic) directly transmit the infection to as much as 20% of the total number affected (Lau et al, 2020).

[8] Scale-free degree distribution was first introduced in the context of citation networks (de Solla Price, 1965). The properties of such networks have been extensively investigated in recent times, although empirical evidence for any social network being scale-free is weak at best (Broido and Clauset, 2019).

Another feature of real-world networks distinguishing them from random graphs which is likely to play a role in shaping the evolution of an epidemic is the existence of strongly clustered neighborhoods, i.e., individuals known to a person would also be extremely likely to be mutually acquainted (Newman, 2003). Typically, such cliques co-occur with short average path-length, which characterize random graphs – a combination which is a hallmark of the so-called "small-world" networks (Watts and Strogatz, 1998). High clustering is expected to decrease the size of the epidemic as transmission from each infected person can only in their local neighborhood; in contrast, spreading can occur extremely rapidly over random networks as a pathogen can travel between any two individuals in only a few hops. Studies on Watts-Strogatz (WS) model networks, which are generated from maximally clustered, regular graphs by rewiring $p$ fraction of the total number of links to randomly chosen individuals in the network, show that even a few random connections (i.e., very low $p$) is sufficient to make the average path length comparable to that of an equivalent random network. Thus, in a clustered social network, the existence of a few connections between neighborhoods that are otherwise far apart will make an outbreak spread in it almost as rapidly as in a random graph. Thus, identification of such links in a specific network and rendering it incapable of transmitting the pathogen (e.g., by quarantining or vaccinating the nodes connected by these links) can provide efficient means of suppressing an epidemic.

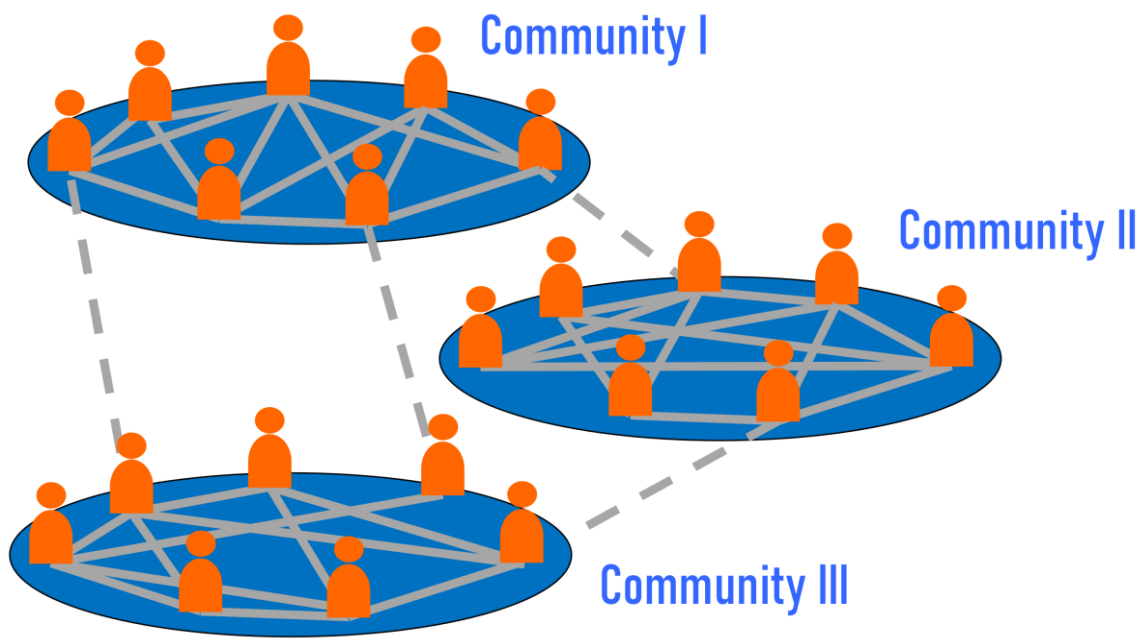

**Figure 5.** Epidemic spreading over a contact network are shaped to a large extent by its mesoscopic structure, which could for example be organized into communities or modules (shown as disks).

It is intriguing that the characteristic features of small-world networks are also seen in networks exhibiting community organization (Pan and Sinha, 2009). Such networks, which capture several empirical features of social networks (Sinha, 2014), are characterized by the existence of modules, subnetworks that have a significantly higher density of connection within members of each module and relatively sparse connectivity between different modules (Fig. 5). It has been suggested by modeling studies that the segregation of different subpopulations (each of which are highly cohesive) can delay the spreading of infection from one module to another, thereby reducing disease burden (Sah et al, 2017). Conversely, it has been shown that such networks can actually promote the long-term persistence of a disease, when individuals lose immunity from prior infection (or vaccination) after a period of time (Jesan et al, 2016). As this is now known to be the case for COVID19, it is relevant to note that the model shows recurrent outbreaks of the disease to be the norm over an extended range of modularity and epidemiological parameters. The reason for this is that as the disease takes a relatively long time to ``jump'' from one community to another, the modular organization allows the epidemic to be long-lasting. We note that evidence of such highly non-uniform spreading behavior has been reported for SARS-CoV-2 infections (Thomas et al, 2020). Thus, by the time the epidemic is about to be extinguished through lack of susceptible, the first individuals to be infected will be just about entering the susceptible compartment once more – which will allow the epidemic to be sustained for longer. The structure of the network allows even relatively small isolated populations to exhibit repeated outbreaks, when a equivalent well-mixed population would have shown extinction of the disease. Thus, the critical community size required for an epidemic to become recurrent can be lowered by the occurrence of mesoscopic structures[9] in the contact network, such as community organization.

**Games**

In the models discussed so far, individuals have almost exclusively been treated as passive recipients of infection. However, to evaluate policy implications of containment measures, one will need to take into account how different individuals will respond to (or adhere to) the specific directives put in place to control an epidemic. A case in point is the use of mass vaccination in an effort to remove the threat of any further outbreaks. Although it may seem intuitive that every individual will choose to vaccinate themselves in order to reduce the risk of infection and disease, and thus, implementing the policy is just a matter of logistics (ensuring that there is enough vaccine and trained personnel to administer the shots), in practice, the situation could be far more complex. Individuals may be tempted to forego the cost or effort involved in getting vaccinated by free-riding on others who have already received the vaccine. This is because herd immunity is a public good, as the immune status conferred by vaccination not only protects the individual receiving it but also benefits the community by screening the unvaccinated individuals from the pathogen. A similar situation arises also for non-pharmaceutical interventions, such as quarantining, travel restrictions, workplace or school closures, and most important from the perspective of COVID19, shelter-in-place (lockdown) mandates. Assuming that the behavior adopted by individuals can be modeled as implementation of strategies adopted by rational agents trying to maximize their personal utility or benefit, based upon information available to them, we can investigate the collective response or adherence using the theory of games.

---

[9] Mesoscopic refers to features that appear neither at the global or macro scale of the network as a whole (such as average path length or diameter) nor at the micro scale of individual nodes (such as their degree) of the network, but at an intermediate level.

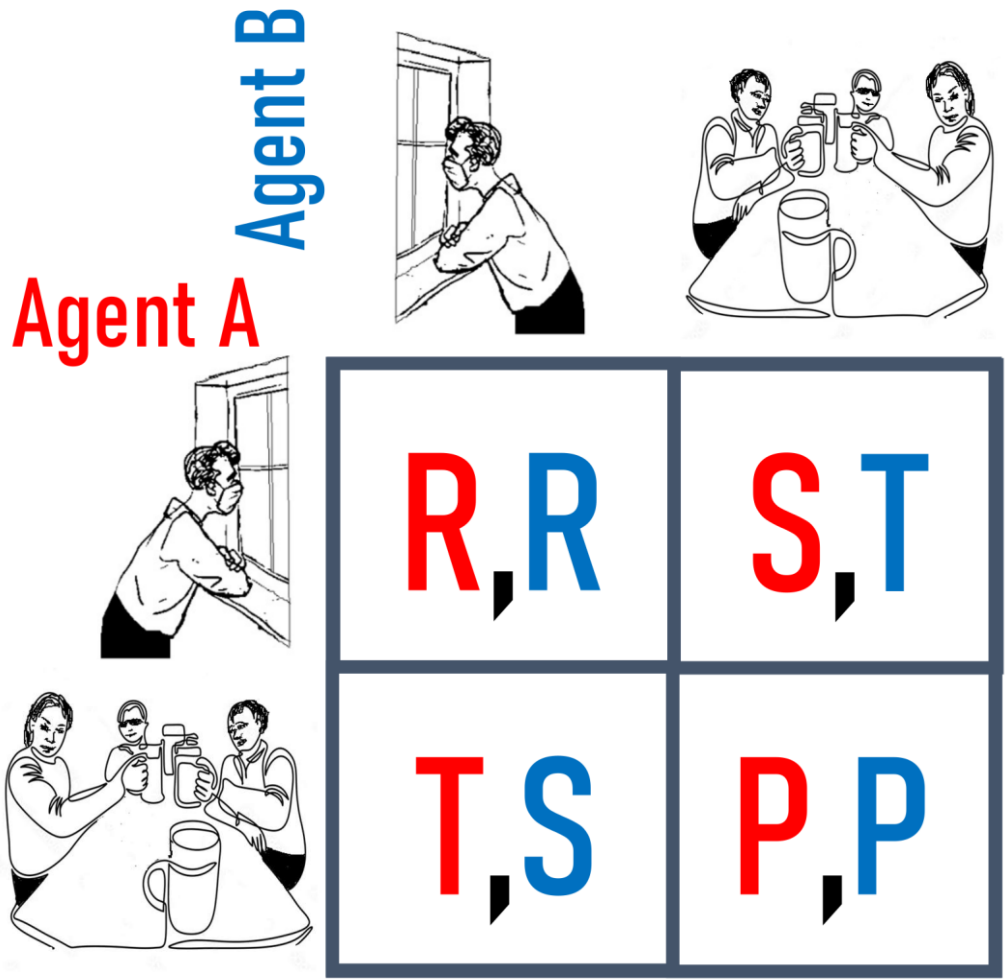

**Figure 6.** Agents deciding on whether to adhere (or not) to restrictions put in place to contain pandemic viewed as a game with a symmetric payoff matrix (Lineart adapted from stockimages).

In this setting, each individual is assumed to receive a payoff depending upon the strategy choice made by all agents including herself. Each agent in turn makes her choice knowing this payoff matrix, and with the understanding that all other agents are also attempting to maximize their utility. In case of collective action aimed at controlling epidemics, higher payoff would correspond to reducing the risk of infection while simultaneously minimizing one's cost in doing so. While, in general, this would be a game involving many agents simultaneously, for simplicity we can focus on the payoff matrix for 2 agents, each having to choose between adherence (cooperation) or not (defection) at each round (Fig. 6). Thus, if both adhere to the epidemic control policy put in place, they each receive a reward payoff R for cooperating, while if both choose not to adhere, the corresponding payoff P is the penalty for mutual defection. If one adheres while the other does not, the former receives the "Sucker's payoff" S, while the payoff T of the latter quantifies the temptation to cheat. For different choices of the relative ordering of the payoffs, situations corresponding to various well-known games such as Prisoners' Dilemma (T>R>P>S) or Hawk-Dove (T>R>S>P) will be realized (Rapoport, 1966). These games illustrate *social dilemmas*, in which actions chosen by agents to maximize their individual payoff can have the paradoxical effect of making the situation worse for everyone (Rapoport and Chammah, 1965). In the case of epidemic control, if every agent believes that everyone else is adhering to the policy imposed (adhering to which involves a personal cost, e.g., restriction of their freedom of movement or being inoculated with a vaccine that may have possible side-effects), they may be tempted to not incur the cost themselves by arguing that as others are all doing it, one person not adhering is unlikely to reduce the efficacy of the policy. This may result in the majority choosing non-adherence resulting in failed implementation of the policy, and consequently increased risk of infection for everyone.

Such arguments have been advanced to explain vaccine hesitancy, a key issue of concern in controlling COVID19 through pharmaceutical means (Solís Arce et al., 2021). Earlier studies have shown that for the success of vaccination programs in reducing the incidence of various childhood diseases may have paradoxically created opportunities for people to consider foregoing it. When a substantially high fraction of the population is already resistant to the disease, the reduced risk of infection makes any perception of even a low risk associated with the vaccine outweigh its benefit (Bauch and Earn, 2004). The myopic focus on only one's immediate payoff can lead individuals to avoid getting vaccinated, creating susceptible pools that provide opportunity for the pathogen to renew its attack and escape eradication. Indeed, the last few years has seen a resurgence in several vaccine-preventable diseases, including measles that had been almost eliminated, in developed economies such as the USA (Phadke et al, 2016). By simulating epidemic propagation on a social network, that is simultaneously used by individuals to estimate risk of infection by obtaining information on disease incidence and number of neighbors vaccinated, to decide on optimal action (whether to get vaccinated or not), it has been observed that the information source for disease prevalence is decisive in shaping public response to vaccination (Sharma et al, 2019). In particular, availability of real-time information about disease incidence in the immediate network neighborhood can lead to higher vaccine acceptance, a potential policy goal for managing COVID19.

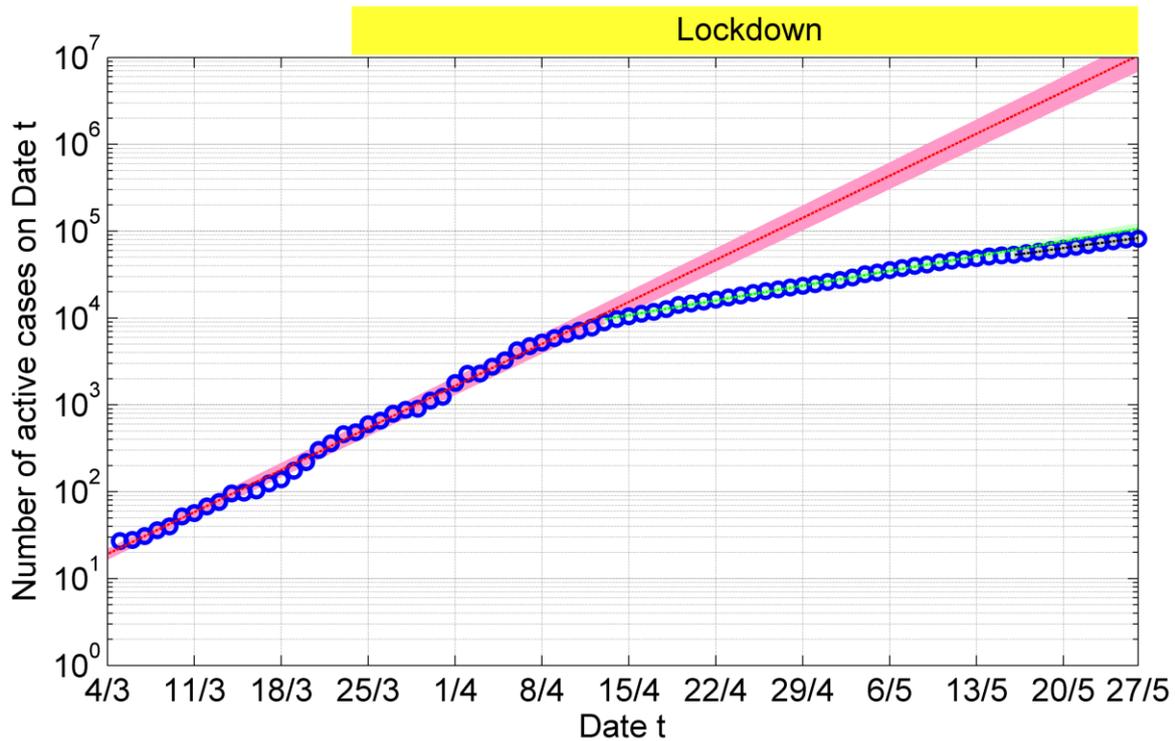

**Figure 7.** The progress of the COVID19 pandemic in India in the period immediately following outbreak. The circles represent the actual incidence while the broken line exhibits a projection of the number of active cases if the initial rate of spreading had continued unchanged. The shaded interval represents the confidence intervals of the projection. The period during which shelter-in-place was imposed on the entire country is indicated by a horizontal bar above the figure.

Similar analysis can also be carried out to understand the effectiveness of non-pharmaceutical interventions. For example, it has been shown using models that social distancing, which reduces disease transmission by decreasing rate of contact between individuals, is most likely to be beneficial for diseases that spread at an intermediate rate, specifically, having basic reproduction number $R_0 \approx 2$ (Reluga,2010). The present pandemic has seen an extensive use of it to buy time (by delaying the progress of the epidemic) to develop a vaccine. However, the most strikingly novel feature of COVID19 management is the widespread imposition of shelter-in-place (lockdown) policies. It has been credited with substantially reducing the burden of the disease, including in India (Fig. 7). However, this measure has a major cost associated with it as it causes widespread psychological, social and economic disruptions (Every-Palmer et al, 2020; Mandel and Veetil, 2020). As the cost of lockdown may be vastly different across socio-economic classes, it is important to understand how the success (or otherwise) of such policies may be tied to the behavior of individuals who may have differing incentives to circumvent such policies. For example, preliminary studies carried out by our group suggest that less than complete adherence to lockdown can lead to multiple resurgent peaks in the disease incidence, resembling the several waves of COVID19 that have occurred so far.

**Challenges and Prospects**

The COVID19 pandemic has been a very instructive episode for the designing of public health policy using mathematical modeling. While the period from the worldwide outbreak in early 2020 has seen a veritable explosion in the number of published modeling studies using an unprecedented range of modeling tools and techniques to study various aspects, it has also revealed the limitations of modeling. While modeling studies have been widely cited to justify one policy or another, it has become clearer how sensitively such results depend on the specific assumptions underlying the modeling. It has also revealed that epidemic progression depends on many different factors, many of which may not even be apparent – a humbling experience for epidemiological modelers who may have imagined that the level of sophistication achieved by the field could finally make it a predictive science. Indeed, there is now greater appreciation that instead of trying to predict the exact time when the next outbreak will occur (which, being a result of many chance factors coming together, is essentially a stochastic event), the true value of modeling lies in providing a testing ground for comparing various remedial measures without having to go through a costly (both in terms of money and lives) trial-and-error process in the field. This has also permeated to the public consciousness, as observed by the fact that in many cases the acceptance of epidemic control policy measures has been contingent on reported results of epidemiological models and the understanding that such modeling has informed the designing of optimal intervention procedures.

Modeling successes and failures during the pandemic have also thrown up important questions that would need to be addressed in the future. For one, the utility of simple vs complex models has been debated with respect to their respective performances in describing the progress of the epidemic. The importance of parameter estimation from noisy data has been appreciated more, as it has been seen that models with large number of parameters do not necessarily give more accurate descriptions of the real-world process. It has also been a puzzle as to why the first wave of the pandemic saw a much lower spreading rate in the poorer countries of South Asia than compared to the develop countries of Europe and North America. The initial facile explanations of this, such as the unfounded speculation about supposedly genetic resistance of South Asians to the pathogen, were patently wrong as shown clearly by the disastrous consequences of the so-called second wave in India. As the disease continues to evolve, with new variants of concerns emerging, increasingly people are wondering about how the epidemic will eventually end. The question of whether mutant strains such as Omicron (i.e., the B.1.1.529 strain of the SARS-CoV-2 virus first detected in Nov 2021 in South Africa) will find it harder to spread because of the partial resistance offered by large sections of the populations being either infected earlier (with different strains) or being vaccinated. Finally, the use of lockdown has emerged as an unprecedented weapon in the arsenal of public health interventions. As its impact had not been known from before, the pandemic has served as an unintended lab for looking at the role of large-scale perturbations upon epidemic spreading, which are however not cost-free. We have to take into account how individuals will weigh the risk of being infected (and potentially dying) against loss of livelihood if we are to use variants of such measures successfully in the future. Possibly one of the most important questions from a policy perspective that future modelers may be called to answer is, if it is possible to have in place a shelter-in-place intervention plan that is socio-economically less damaging while having the same effectiveness as the lockdown that has been implemented during the current pandemic. The results of such studies promise to usher in a new age of modeling-driven public health policy design.

**Acknowledgments**

The author would like to acknowledge extensive discussions at different times over the past several years with T Jesan, Chandrashekar Kuyyamudi, Gautam I Menon, Shakti N Menon, Raj K Pan, V Sasidevan, Anupama Sharma and Somdatta Sinha. Generous funding support is acknowledged for the Center of Excellence in Complex Systems and Data Science (CoE-CSDS) from the Department of Atomic Energy, Government of India.